\documentclass[12pt]{article} % 10pt is ignored!

\usepackage{epsfig}
\usepackage{graphicx}% Include figure files
\usepackage{amsmath}
\usepackage{amssymb}

\begin{document}

\title{On the effect of ocean tides and tesseral harmonics on spacecraft flybys of the Earth}

% The list of authors, and the short list which is used in the headers.
% If you need two or more lines of authors, add an extra line using \newauthor
\author{L. Acedo\thanks{E-mail: luiacrod@imm.upv.es}\\
Instituto Universitario de Matem\'atica Multidisciplinar,\\
Building 8G, $2^{\mathrm{o}}$ Floor, Camino de Vera,\\
Universitat Polit$\grave{\mbox{e}}$cnica de Val$\grave{\mbox{e}}$ncia,\\
Valencia, Spain\\
}

\maketitle

\begin{abstract}
The so-called flyby anomaly has encouraged several authors to analyze in detail the 
minor perturbative contributions to the trajectory of spacecraft performing a
flyby manoeuvre. This anomaly consist of an unexplained increase or decrease of the asymptotic
velocity of the spacecraft after a flyby of the Earth in the range of a few mm per second.
Some order of magnitude estimations have been performed in recent years to dismiss many
possible conventional effects as the source of such an anomaly but no explanation has been found yet.
In this paper we perform a study of the perturbation induced by ocean tides in a flybying spacecraft
by considering the time dependence of the location of the high tide as the Moon follows its orbit.
We show that this effect implies a change of the spacecraft velocity of a few micrometers per second.

We also consider the coupling of tesseral harmonics inhomogeneities and the rotation of the Earth
and its impact of the spacecraft outgoing velocity. Significant corrections to the observed asymptotic velocities are found in this case but neither their sign nor their magnitude coincide with the anomalies. So, we can also rule this out as a conventional explanation.
\end{abstract}



%%%%%%%%%%%%%%%%%%%%%%%%%%%%%%%%%%%%%%%%%%%%%%%%%%

%%%%%%%%%%%%%%%%% BODY OF PAPER %%%%%%%%%%%%%%%%%%

\section{Introduction}
\label{sec:1}
For the most part of human history astronomy has been a science based upon observations of celestial
bodies but with the emergence of astrodynamics in the mid-twentieth century this situation has changed
\cite{BMW}. Nowadays it is possible to perform accurate measurements of spacecraft trajectories and to obtain
direct information for the planets and moons of the Solar system. Moreover, the deployment of retrorreflectors in the Moon's surface by the Apollo missions has allowed the development of the Lunar
Laser Ranging technique by which the Moon's location in space is determined with unprecedented accuracy
\cite{Chapront,Dickey1994,Williams1996,Williams2004}.

These new tools mark the beginning of an era of high-precision astronomy and astrodynamics in which 
effects, previously below the level of the accuracy of observations, are now disclosed with increasing
frequency. As a canonical example we should cite the history of the Pioneer anomaly and its recent 
solution in terms of thermal emission by the spacecraft \cite{TuryshevReview,PioneerPRL,Rievers2011,Bertolami2008,Bertolami2010,Bertolami2012}. It is well-known that a discrepancy between the modelled and the predicted Doppler data has been noticed in both the Pioneer 10 and Pioneer 11 spacecraft. This effect was interpreted as a constant acceleration of $a_P=(8.74\pm 1.33) \times 10^{-8} $ cm$/$s$^2$ directed towards the Sun \cite{LPDSolarSystem,Anderson2002}. For many researchers, this minute discrepancy suggested that
new physics was operating and it stimulated many proposals beyond standard General Relativity but several studies showed that the planets cannot be influenced by an acceleration of similar magnitude
\cite{Tangen2007,FiengaPioneer,IorioPioneer2006,IorioPioneer2007,IorioPioneer2010}. Later on,
the retrieval of the entire telemetry data at the Jet Propulsion Laboratory (dating back to the early stages of the mission in the seventies of the past century) was instrumental in the elucidation of this anomaly because it revealed a trend in the extra acceleration. This trend was consistent with a recoil acceleration  arising from the thermal anisotropic emission of the heat delivered by the Radioisotope Thermoelectric Generator (RTG) as well as the electric instruments onboard \cite{PioneerPRL,IorioIJMPDReview}. 

Approximately at the same time, it was also found that fitting the post-encounter residuals of the spacecraft trajectories performing flybys of the Earth leads also to unexplained discrepancies \cite{Anderson2008}. In terms
of velocities, these discrepancies correspond to an increase or a decrease of the asymptotic outgoing spacecraft velocity with respect to the ingoing velocity that cannot be fitted with the orbit determination programs. The difference amounts to a few mm per second and it is also found in the ranging data. Some attempts have been done to look for an explanation based upon conventional physical effects but to no avail: (i) L\"ammerzahl et al. considered the order of magnitude of atmospheric drag, ocean and solid Earth tides, charging of the spacecraft, magnetic moment, Earth albedo, Solar wind and spin-rotation coupling and they concluded that they were very small to account for the anomaly \cite{LPDSolarSystem} (ii) Iorio studied the effect of General Relativity on Hyperbolic Orbits considering both gravitomagnetic and gravitoelectric effects but the maximum deviations are five order of magnitude below the detected flyby anomaly \cite{IorioSRE2009} (iii) Thermal effects similar to that responsible of the Pioneer anomaly were considered by Rievers and 
L\"ammerzahl showing that they cannot be responsible of the flyby anomaly \cite{Rievers2011} (iv) Atchison et al. studied the Lorentz acceleration of a charged spacecraft but they conclude that it is unlikely that they could
completely explain away the anomalies in this context \cite{Atchison} (v) Hackmann and L\"ammerzahl have also analyzed the Lense-Thirring effect for hyperbolic orbits with similar negative results \cite{Hackmann}.

The claims for an origin of the flyby anomaly beyond standard physics started with the seminal work of
Anderson et al. \cite{Anderson2008}, in which the anomalies of six flybys of the Earth from 1990 to 2005 were discussed. In this work a phenomenological formula was proposed as a fit for the anomalous energy changes. Anderson et al. \cite{Anderson2008} claim that the energy change is proportional to the variation of the cosine of the declinations for the incoming and outgoing velocity vectors of the spacecraft. This correlation suggest that a relation with Earth's rotation is operating and the authors referred to an enhanced Lense-Thirring effect. Moreover, another anticorrelation with the sign of the azimuthal velocity at perigee has recently been analyzed in the context of an extended Whitehead's model
of gravity \cite{Acedo2016}. This means that the sign of the anomaly is positive for flybys in which the azimuthal velocity at perigee is opposite to Earth's rotation and viceversa. 

For a purely empirical point of view, a correlation with the altitude of the perigee (the effect is larger for  smaller altitude) is also found \cite{Jouannic}. 
Another obvious correlation is found by simple inspection of the results listed by Anderson et al. \cite{Anderson2008}: the geocentric latitude of the perigee and the sign of the anomaly seem to be related (being positive for flybys with a perigee at the Northern hemisphere). These unexpected correlations make difficult to find a simple explanation as a systematic effect arising for an unmodelled classical source.

Intrigued by this mysterious anomaly several researchers have looked for models beyond standard physics:
Adler has studied the possibility of an halo of dark matter surrounding the Earth and its effect on 
spacecraft flybys \cite{Adler2010,Adler2011}; other approaches imply modifications of Newtonian gravitation or General Relativity
more or less well-motivated \cite{Nyambuya2008, Lewis2009,Hafele2009,Acedo2014,Varieschi2014,Pinheiro2014,Acedo2015,Wilhelm2015,Pinheiro2016,Bertolami2016}. But we have still no convincing explanation of the phenomenon of flyby anomalies. Occam's razor dictates that all conventional explanations should be carefully analyzed and 
dismissed before claiming that new physics is necessary in this case. The objective of this paper is to
perform a quantitative estimation of the energy transfer from tides and tesseral harmonics to a spacecraft performing a flyby of the Earth \cite{Torge1991}. As the location of the high tide changes with time and the tesseral inhomogeneities follow the Earth in its rotation both effects create a time-dependent gravitational potential which causes small energy changes in the spacecraft. However, we will show that these are not sufficient to explain the observed anomalies.

A similar contribution by the tesseral harmonics is found to be significant but, on the other hand, insufficient to explain the anomalies.

\section{Ocean tides and energy transfer to spacecraft during a flyby}
\label{sec:2}
The accurate study of ocean tides is a classic problem in geodesy which starts with the system of Laplace's tidal equations and the disturbing potential of the Moon and the Sun \cite{Torge1991}. On the other hand, the different topographies of coastlines and shelf areas induce oscillations and greatly complicate the problem of finding the local height of the tide. This results in a complex pattern of cotidal lines and amphidromic points, i. e., the points of zero amplitude for the principal harmonic constituent of the tide where the cotidal lines met. 

As we are interested in finding a quantitative upper bound for the effect of tides upon spacecraft perfoming flyby manoeuvres around the Earth some simplifications are recommendable. We will assume that
the Earth is a spherical planet with a constant depth global ocean in the absent of tides. These tides
are the consequence of the Moon's gravitational pull and the resulting sea profile level can be approximated by a Jacobi or scalene ellipsoid in which the maximum height of the tide takes place directly in the intersection of the Moon's position vector and the Eart's surface, i. e., the Earth's location in which the Moon is at its zenith (and also in the antipodes of this place). 

\begin{figure}
	\includegraphics[width=\columnwidth]{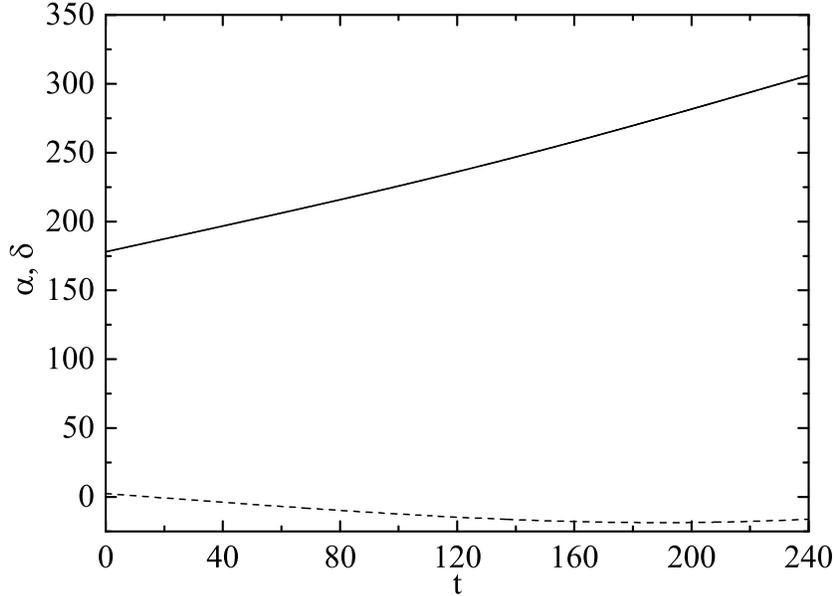}
    \caption{The right ascension (solid line) and the declination (dotted line) of the Moon in sexagesimal degrees from January 18th at $00$:$00$ UTC. Time is measured in hours.}
    \label{fig1}
\end{figure}

We must notice that the location of this point with respect to the fixed stars changes as the Moon follows its orbit. The total gravitational potential of the Earth including the effect of the ocean tide is given by:
\begin{equation}
\label{utide}
U(\mathbf{r})=-\displaystyle\frac{G M}{r}+\displaystyle\frac{G}{2 r^3} \, \left(C - A\right)\, \left( 3 \cos^2 \theta-1\right)\; ,
\end{equation}
where $M$ is the mass of the Earth, $r$ the distance from the center of the Earth to the point of interest (the spacecraft in our case), $\theta$ is the angle among the spacecraft and the Moon's position vectors and $C-A$ is the difference among the moment of inertia with respect to the axis corresponding to the direction of the Moon and another axis perpendicular to it. We have that:
\begin{equation}
\label{CA}
C-A=\displaystyle\frac{M}{5} \left[ R^2_{\mbox{max}}-R^2_{\mbox{min}} \right] \simeq \displaystyle\frac{2}{5} \, M \, R_{\mbox{geo}} \, h_{\mbox{tide}} \; ,
\end{equation}
where $R_{\mbox{max}}$, $R_{\mbox{min}}$ are the maximum and minimum Earth radius taken into account the
height of the tide, $R_{\mbox{geo}}$ is the average radius of the Earth including the average ocean depth and $h_{\mbox{tide}}$ is the maximum height of the tide. We must also take into account that $\cos \theta= \mathbf{\hat{r}} \cdot  \mathbf{\hat{R}}$, where $\mathbf{\hat{r}}$ and $\mathbf{\hat{R}}$ are the unit vectors in the direction of the spacecraft and the Moon, respectively. Then, from Eqs. (\ref{utide}) and (\ref{CA}) we get:
\begin{equation}
\label{utidef}
U(\mathbf{r})=-\displaystyle\frac{G M}{r}+\displaystyle\frac{G M}{5 r^3}\, h_{\mbox{tide}}\,  R_{\mbox{geo}} \,
\left(3 \cos^2 \theta -1 \right)\; \; .
\end{equation}
Notice that this is a time-dependent potential because it depends on the unit position vector of the Moon, $\mathbf{\hat{R}}(t)$. The partial derivative with respect to time is then given as follows:
\begin{equation}
\label{Utidedt}
\displaystyle\frac{\partial U}{\partial t}=\displaystyle\frac{\mu}{R_{\mbox{geo}}} \, \left( \displaystyle\frac{R_{\mbox{geo}}}{r} \right)^3 \displaystyle\frac{6 h_{\mbox{tide}}}{5 R_{\mbox{geo}}}
\,\mathbf{\hat{r}}\cdot \mathbf{\hat{R}} \left( \mathbf{\hat{r}} \cdot \displaystyle\frac{\partial
\mathbf{\hat{R}}}{\partial t} \right) \; ,
\end{equation}
with $\mu= G M=398675.0573$, km$^3$/s$^2$, as the value of the Earth's mass constant and $R_{\mbox{geo}}=6371$ km, its radius. An upper bound for the tide's height is $h_{\mbox{tide}}=10$ m (as it is well-known the maximum ocean tides in Earth are found in the Bay of Fundy with extremes of $16$ m as a consequence of the special geography of the region \cite{Desplanque}). We will now calculate an estimation for
the partial derivative of the potential during the NEAR flyby of January 23rd, 1998. 
The right ascension and the declination of the Moon in a period of ten days starting in January 18th is plotted in Fig. \ref{fig1}. The Moon's position vector in celestial equatorial coordinates is then
obtained as a function of time:
\begin{equation}
\mathbf{\hat{R}}=\sin \delta \, \cos\alpha \, \boldsymbol{\hat{\textbf{\i}}} +\sin\delta\, \sin \alpha\, \boldsymbol{\hat{\textbf{\j}}}+
\cos\delta \, \mathbf{\hat{k}} \; .
\end{equation} 
From Eq. (\ref{Utidedt}) we can calculate the partial derivative of the potential at every instant
along the spacecraft's trajectory. As this trajectory we can use the approximate osculating keplerian
orbit at perigee with eccentricity: $\epsilon=1.81352$, semi-major axis: $a=-8494.87$ km, magnitude of
the velocity, $V_p=12.7401$ km$/$s, right ascension for the perigee, $\alpha_P=280.42$ and declination,
$\delta_P=33$ sexagesimal degrees. Alternatively, we can use the interpolated trajectory from the ephemeris for NEAR. 

\begin{figure}
	\includegraphics[width=\columnwidth]{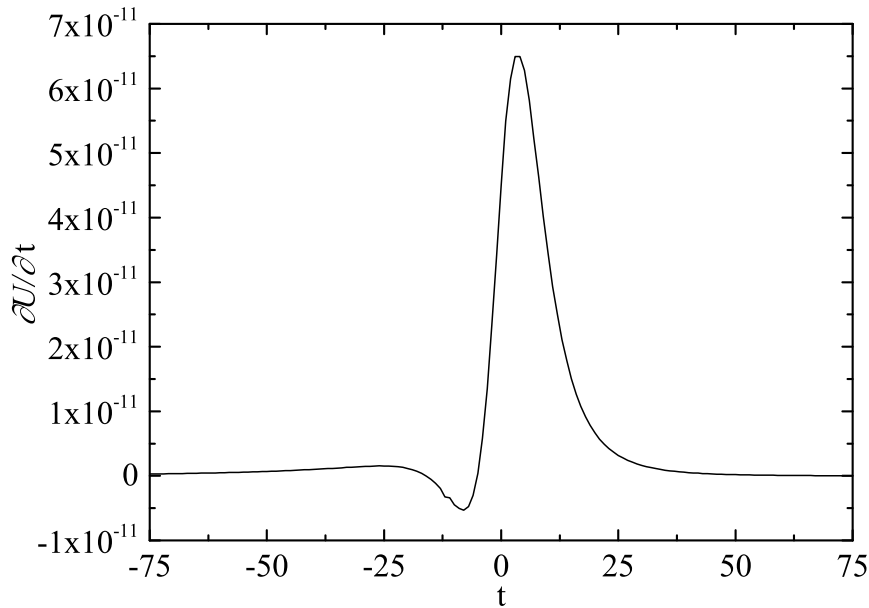}
    \caption{Partial derivative of the potential (per unit mass) arising from the moving tide along the trajectory of
    the NEAR spacecraft in km$^2/$s$^3$. Time is measured in minutes and the point of closest approach corresponds to $t=0$.}
    \label{fig2}
\end{figure}
In Fig. \ref{fig2} we show the partial derivative of the potential as given by Eq. (\ref{Utidedt}) per unit mass as a function of time. The time derivative of the Moon's position vector is obtained from the
Moon's ephemerides as plotted in Fig. (\ref{fig1}).

\begin{figure}
	\includegraphics[width=\columnwidth]{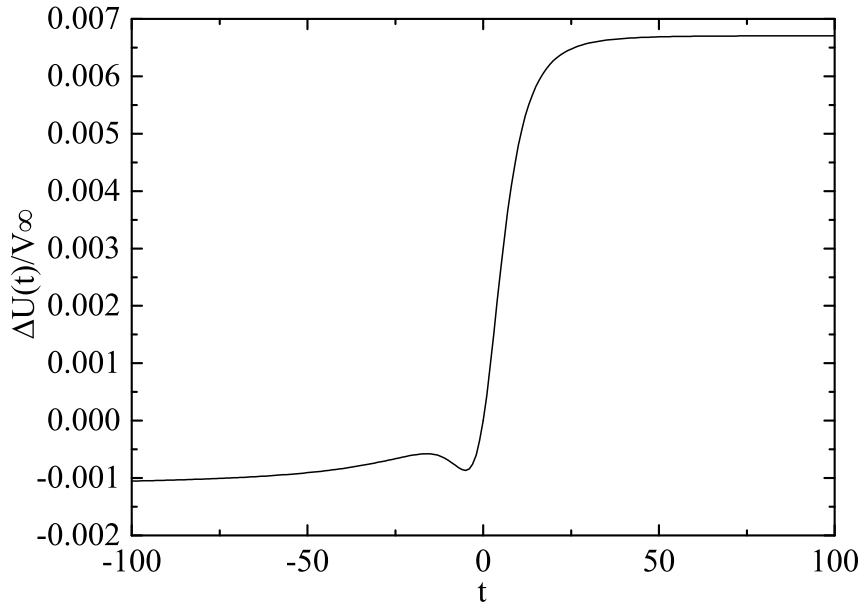}
    \caption{Variation of the NEAR's spacecraft asymptotic velocity as a consequence of the perturbing effect of the ocean tide. The vertical axis gives the velocity change in mm per second. The horizontal axis is the time in minutes with the point of closest approach as reference.}
     \label{fig3}
\end{figure}

A result of classical physics for time-dependent potentials identifies the total derivative with the
partial derivative and, consequently, the variation in total energy along the spacecraft trajectory can
be calculated from the integral:
\begin{equation}
\label{Uint}
\Delta U=\displaystyle\int_{t_i}^{t_o}\, d t \displaystyle\frac{\partial U}{\partial t}\; ,
\end{equation}
where $t_i$ and $t_o$ denote the, adequately chosen, incoming and outgoing times for the flyby manoeuvre. It is more convenient to give the change in asymptotic velocity for the osculating orbit at
each point of the real orbit as $\Delta V_{\infty}=\Delta U/V_{\infty}$. In Fig. \ref{fig3} we have plotted the result of this integration for the NEAR flyby and a period of time starting 100 minutes before the perigee and ending 100 minutes after the perigee. 
The total velocity change is, approximately, $\Delta V_{\infty} \simeq 0.0078$ mm$/$s and, obviously, it is too small to account for the flyby anomaly which was evaluated as $13.46$ mm$/$s by Anderson et al. in this particular flyby. In the next section we will consider the effect of another time-dependent potential: the one generated by the tesseral harmonics as the Earth rotates around its axis.
\section{Tesseral harmonics contribution to perturbations of spacecraft orbits}
\label{sec:3}
Local inhomogeneities of the Earth's gravitational field can be modelled in terms of an expansion in
spherial harmonics. The resulting geopotential model is given by the following series expansion:
\begin{equation}
\label{geomodel}
\begin{array}{rcl}
U(r,\theta,\lambda)&=&-\displaystyle\frac{\mu}{r} \left[1+\displaystyle\sum_{n=2}^\infty\, 
\displaystyle\sum_{m=0}^n \, \left( \displaystyle\frac{R}{r} \right)^n \, P_{nm}\left(\cos \theta\right) \right. \\
\noalign{\smallskip}
& &\left. \left\lbrace C_{nm} \cos\left( m \lambda \right)+S_{nm} \sin \left( m \lambda\right) \right\rbrace \right]\; ,
\end{array}
\end{equation}
where $\theta$ is the polar angle, $\lambda$ is the geocentric latitude and $R$ km is a normalization
quantity giving a measure of the Earth's radius. The functions $P_{nm}(x)$, $n,m=0$, $1$, $\ldots$ are
the associated Legendre polynomials:
\begin{equation}
\label{legendre}
P_{nm}(x)=\displaystyle\frac{1}{2^n n!} \left(1-x^2\right)^{m/2} \, \displaystyle\frac{d^{n+m}}{d x^{n+m}}\, \left( x^2-1\right)^m\; .
\end{equation}
From Eqs. (\ref{geomodel}) and (\ref{legendre}) we have that the lowest order correction to the Newtonian potential of a perfectly spherical planet is:
\begin{equation}
\label{UJ2}
U(r,\theta,\lambda)=-\displaystyle\frac{\mu}{r}\left[1+\left(\displaystyle\frac{R}{r} \right)^2
\displaystyle\frac{1}{2} \left( 3 \cos^2 \theta-1\right) C_{20}+\ldots\right]\; ,
\end{equation}
and here we identify $C_{20}$ as the scaled zonal term of order two which takes the value, $C_{20}=J_2=1.0826 \times 10^{-3}$ for the reference radius $R=6378.1363$ km. In general, the terms in 
Eq. (\ref{geomodel}) for $m=0$ are called zonal harmonics and these depend only on the polar coordinate.
The terms with $m\neq 0$ depend on the latitude, $\lambda$, as well and they are referred to as tesseral
harmonics. 

The precision and number of coefficients known in the expansion of Eq. (\ref{geomodel}) dramatically improved in the last three decades of the past century. The EGM96 model, now updated to EGM2008, is 
still used in many studies as it provides reasonable accuracy for the terms up to order $n=360$, $m=360$. We must also take into account that the coefficients in the EGM96 tables are related to the
ones used in Eq. (\ref{geomodel}) by the expression \cite{Vallado}:
\begin{equation}
C_{nm}=\left[ \displaystyle\frac{(n-m) ! k (2 n +1)}{(n+m) !} \right]^{1/2} \, \bar{C}_{nm}\; ,
\end{equation}
where $k=1$ for $m=0$ and $k=2$ for $m\neq 0$ and $\bar{C}_{nm}$ are the tabulated coefficients. The 
same expression holds for $S_{nm}$. 

The zonal part of the potential is conservative and it does not contribute to the change in asymptotic
energy for the spacecraft. On the other hand, we must notice that the latitude of the vertical of a
star, fixed with respect to the celestial equatorial system of reference, changes as the Earth rotates around its axis. Consequently, in the celestial system of reference the potential in Eq. (\ref{geomodel}) depends explicitly on time. For a spacecraft performing a flyby in which the latitude
of the vertical of the closest approach is $\lambda_{\mbox{p}}$ and the right ascension of the spacecraft at that instant is $\alpha_{\mbox{p}}$ we have:

\begin{equation}
\label{lambda}
\lambda(t)=\lambda_{\mbox{p}}-\alpha_{\mbox{p}}+\alpha(t)-\Omega \, t\; ,
\end{equation}

where $\Omega=2 \pi/86400$ rad$/$s is the angular velocity of the Earth's rotation around its axis.
We have also taken into account that both the geocentric latitude and the right ascension of the 
spacecraft are measured eastward and the Earth rotates in the same direction.

\begin{figure}
	\includegraphics[width=\columnwidth]{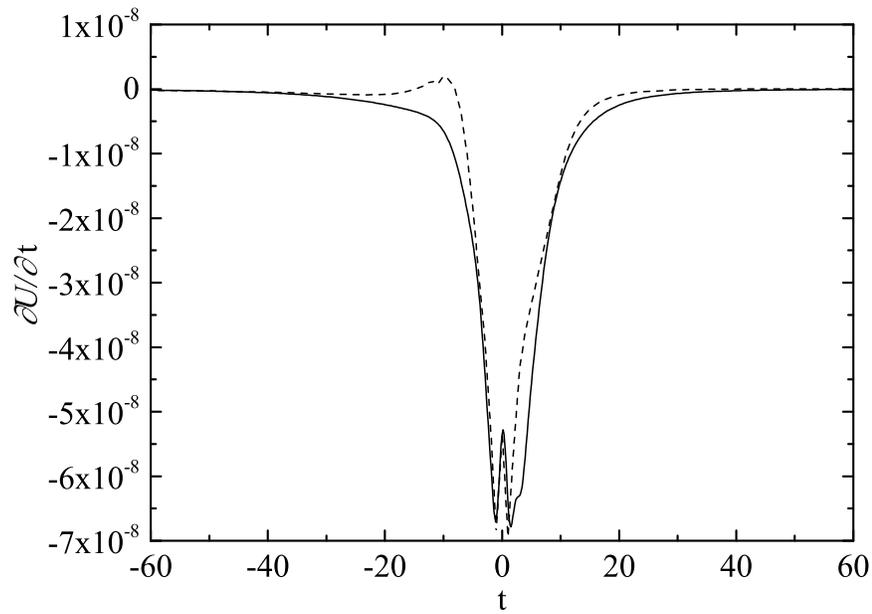}
    \caption{Partial derivative of the potential (per unit mass) obtained from Eq. (\protect\ref{DUdt})
    for  the NEAR flyby in km$^2/$s$^3$ vs time in minutes. The solid line corresponds to the real trajectory and the dotted line to the osculating orbit at perigee (as given in the previous section). A total of $N=360$ terms were considered in the sum as given by the EGM96 geopotential model.}
    \label{fig4}
\end{figure}

From Eqs. (\ref{geomodel}) and (\ref{lambda}) we now obtain:

\begin{equation}
\label{DUdt}
\begin{array}{rcl}
\displaystyle\frac{\partial U}{\partial t}&=&\displaystyle\frac{\mu}{r} \, \Omega \, 
\displaystyle\sum_{n=2}^N\, 
\displaystyle\sum_{m=0}^n \, m \, \left( \displaystyle\frac{R}{r} \right)^n \, P_{nm}\left(\cos \theta\right) . \\
\noalign{\smallskip}
& & \left\lbrace -C_{nm} \sin\left( m \lambda(t) \right)+S_{nm} \cos \left( m \lambda\right) \right\rbrace \; ,
\end{array}
\end{equation}
where $N$ is the number of terms considered in the geopotential model we use. In Fig. \ref{fig4} we
have plotted the results for the partial derivative in Eq. (\ref{DUdt}) in the case of the NEAR flyby and for the EGM96 model which contains $360 \times 360$ terms. We notice that the result is three order of magnitude larger than the one corresponding to the effect of tides as shown in Fig. \ref{fig2}.

\begin{figure}
	\includegraphics[width=\columnwidth]{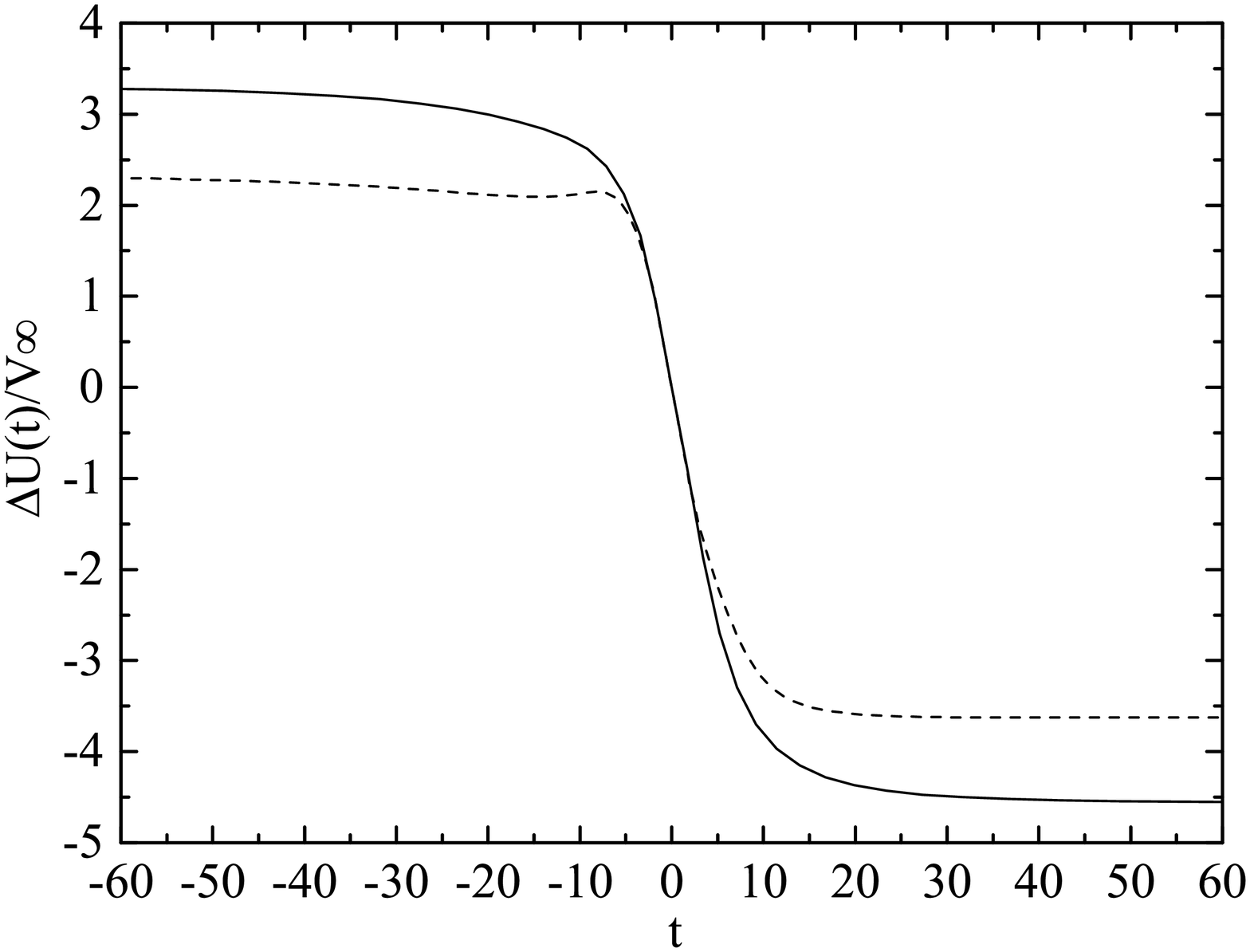}
    \caption{The same as Fig. \protect\ref{fig3} but for the tesseral harmonics. The solid (dotted) line correspond to the integration along the real trajectory  (osculating orbit at perigee). Notice that we obtain now a decrease in the range of a few mm$/$s.
}
    \label{fig5}
\end{figure}

The integration over time gives us the energy change of the spacecraft as a consequence of the time-dependent interaction with the tesseral harmonics. The result is displayed in Fig. \ref{fig5}.
 
\begin{figure}
	\includegraphics[width=\columnwidth]{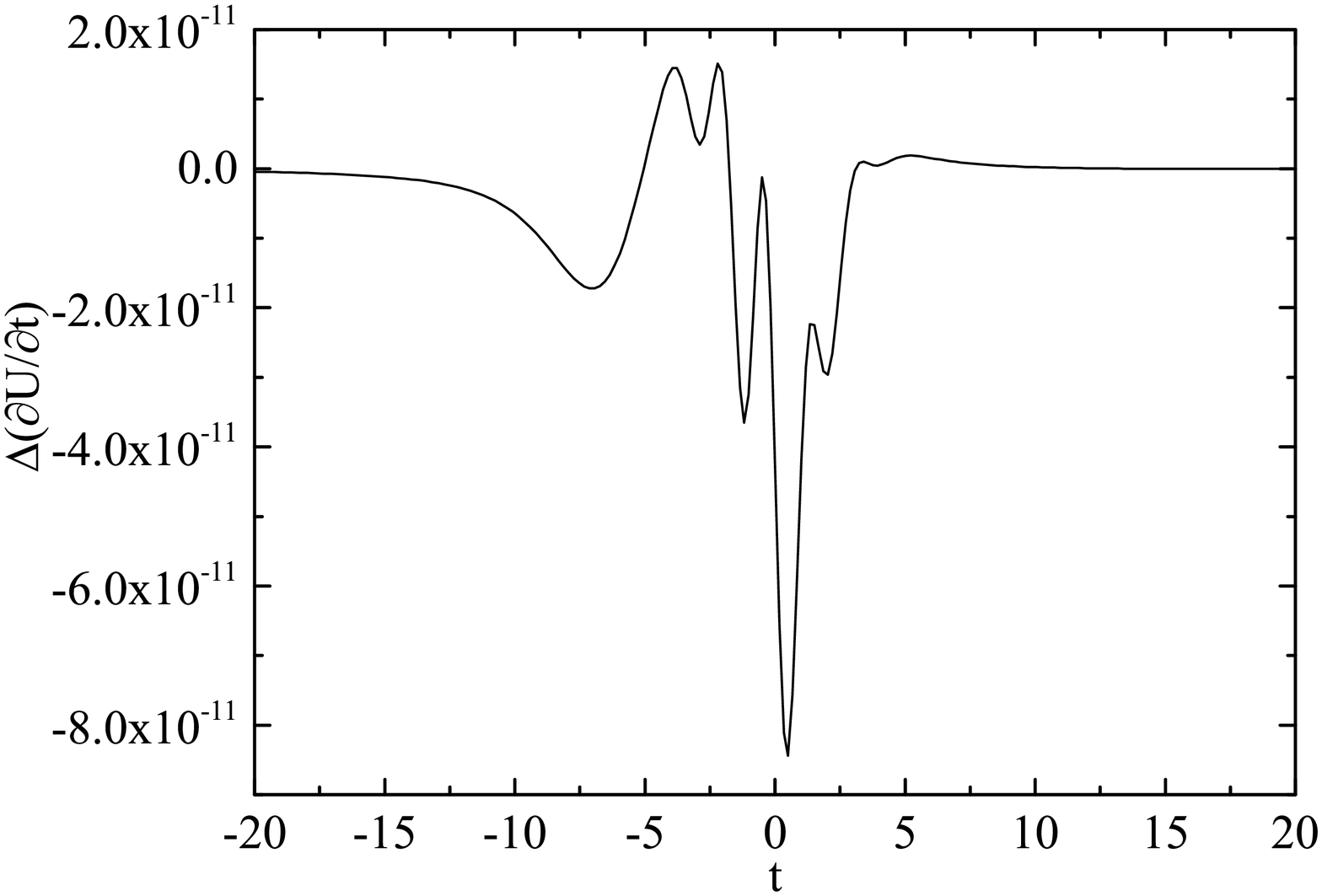}
    \caption{The same as Fig. \protect\ref{fig4} but for the error arising from the uncertainty in the coefficients.}
    \label{fig6}
\end{figure}

From this figure we find that $\Delta V_{\infty}=-5.953$ mm$/$s. This value is comparable with the observed anomaly for the NEAR flyby but its sign is opposite. The error in this value arising from
the uncertainty of the geopotential model's coefficients is small. The estimated error in the derivative $\partial U/\partial t$ is shown in Fig. \ref{fig6} for the same NEAR flyby and we conclude that it can 
be safely ignored because it is, at most, one thousandth of the values of this derivative.

In the table \ref{tab1} we show the results for the amount of asymptotic velocity change as given from 
the energy transfer of the tesseral harmonics. We also list the observed velocity 
change as a comparison.

\begin{table}
\begin{center}
\begin{tabular}{lcccr} % four columns, alignment for each
		\hline
Spacecraft Flyby & Date & Perigee (km) & $\Delta V_{\infty}$ & $\Delta V_{\infty}$ (obs)\\
		\hline
		NEAR & 1/23/1998 & 539 &  -5.95 & 13.46  \\
%\noalign{\smallskip}		
		GALILEO I & 12/8/1990 & 960 & 0.53 & 3.92  \\
%\noalign{\smallskip}		
		GALILEO II & 12/8/1992 & 303 &  -0.76 & -4.6  \\
%\noalign{\smallskip}
   		CASSINI & 8/18/1999 & 1175 & -0.25 & -2   \\
%\noalign{\smallskip}
        ROSETTA & 3/4/2005 & 1956 & -0.64 & 1.8  \\
%\noalign{\smallskip}
        ROSETTA II & 13/11/2007 & 5322 & -0.39 & 0\\
%\noalign{\smallskip}
        ROSETTA III & 13/11/2009 & 2483 &  3.36 & 0 \\
%\noalign{\smallskip}
        MESSENGER  & 8/2/2005 & 2347 & -4.281 & 0.02  \\
%\noalign{\smallskip}
        JUNO & 9/10/2013 & 559 & 0.383  & 0    \\          		
		\hline
	\end{tabular}
\end{center}	
\caption{Observed velocity changes in mm$/$s vs the predictions for the energy transfer arising from the tesseral harmonics' time-dependent potential also in mm$/$s. We give the results for nine flybys performed between 1998 and 2013. The altitude of the perigee in kms is also shown.\label{tab1}}
\end{table}

So, it is clear that the flyby anomaly cannot be explained at a consequence of the energy transfer
induced by the time-dependent tesseral potential. Moreover, the signs of the effect are different
from those of the detected anomaly in most cases and this means that, if it is not accounted for or it is miscalculated by the orbit determination program (ODP), we will have a significant modification of the experimental results but the anomaly would still persist. 

\section{Conclusions}

High-accuracy monitoring of spacecraft flybys requires the consideration of many classical effects
in order to discard them as being significant, within the precision we are obtaining with the Doppler
tracking measurements, or to incorporate them into the orbital model. The effects that
have been listed in the literature as possible sources of noticeable perturbations are: (i) the atmospheric friction for spacecraft traveling through the termosphere, (ii) the gravitational interaction of ocean and solid tides on the spacecraft, (iii) the charge and magnetic moment of the spacecraft, (iv) the pressure caused by the Earth's albedo and solar wind, (v) corrections provided
by General Relativity or (vi) the effect of Earth's oblateness and inhomogeneities computed through
the zonal and tesseral harmonics. Some of these effects have been estimated but the importance of the
problem of the flyby anomalies demands that accurate calculations should be done for each of them
\cite{LPDSolarSystem}. 

In this paper we have calculated a bound on the perturbation induced by ocean tides on a spacecraft flyby around the Earth. This is, at least, three orders of magnitude below the velocity change 
deduced from the Doppler shift residuals. The rotating Earth also generates a time-dependent tesseral
potential on any approaching spacecraft and, by using the EGM96 geopotential model, we have calculated
the resulting energy transfer for each flyby since the Galileo flyby of 1990. We conclude that its
contribution to the variation in the magnitude of the spacecraft's velocity vector is below one millimeter per second in most cases. Nevertheless, statistically significant contributions within the error bars are found for the NEAR, Messenger and Rosetta II flybys. These should be taken into account in the computation of the total flyby anomaly but, on the other hand, we have found that they cannot explain the anomalies and this diminish the number of options for a purely classical explanation using
an overlooked effect. 

From the analysis of several flybys in the period from 1990 to 2005 we know that these anomalies are
evident in the Doppler and ranging data but they are also puzzling for several reasons: the correlation 
among the flyby anomaly sign and the azimuthal velocity and latitude at perigee \cite{Acedo2016}, the lack of detection of any anomaly in the low altitude Juno flyby of Earth in 2013 \cite{Jouannic}, similar null results for the Rosetta II and Rosetta III flybys or the manifestation of the anomalies both in the ranging and Doppler data and also with different orbit determination programs at NASA and ESA \cite{Anderson2008}.

This is in contrast with the case of the Pioneer anomaly whose systematic origin was clear once the whole
data record was analyzed \cite{PioneerPRL}. No such a clear pattern has ever been found for the flyby anomaly and the classical effects studied to date, including the contributions from ocean tides and tesseral harmonics discussed in this paper, are lacking in providing a satisfactory explanation. For these
reasons, we cannot exclude that the origin of the flyby anomaly could come from effects beyond standard
physics.

We hope that the Juno mission to Jupiter \cite{JunoMissionI,JunoMissionII,JunoMissionIII} could help to obtain new data as this spacecraft is scheduled to perform many low altitude flybys over the top clouds of the planet (roughly at 5000 kms). If this phenomenon is real, and it is not the result of a miscalculation, it should appear more clearly in this case as Jupiter's gravitational field and angular momentum is much larger than that of the Earth.

\section*{Acknowledgements}

Ll. Bel is acknowledged for some useful comments on this paper. NASA's Jet Propulsion Laboratory and J. Giorgini are also ackowledged for providing all the ephemerides of this work through the on-line Horizon system and some helpful comments. 
%%%%%%%%%%%%%%%%%%%%%%%%%%%%%%%%%%%%%%%%%%%%%%%%%%



\bibliographystyle{plain}
\bibliography{acedobiblio} % if your bibtex file is called example.bib

\end{document}